# One micron length scale controls kinetic stability of low energy glasses

Kenneth L. Kearns, M. D. Ediger*

*Department of Chemistry, University of Wisconsin-Madison, Madison Wisconsin 53706 USA*

Heiko Huth, Christoph Schick

*Institute of Physics, University of Rostock, Rostock, 18051 Germany*

* to whom correspondence should be addressed: ediger@chem.wisc.edu

Abstract:

AC nanocalorimetry was used to measure the reversing heat capacity $C_p$ of low energy indomethacin glasses as they isothermally transform into the supercooled liquid. As the film thickness increases from 75 to 600 nm, the transformation time increases by more than an order of magnitude, consistent with a surface-initiated transformation mechanism. Eventually, the transformation time becomes constant for films between 1.4 and 30 μm indicating a distinct bulk transformation pathway. The observation of size-dependent transformation kinetics for glass samples approaching 1 μm is unprecedented. We interpret the crossover in thickness dependence at 1 μm to signify the average distance between transformation initiation sites in the bulk low energy glass.

TOC figure:

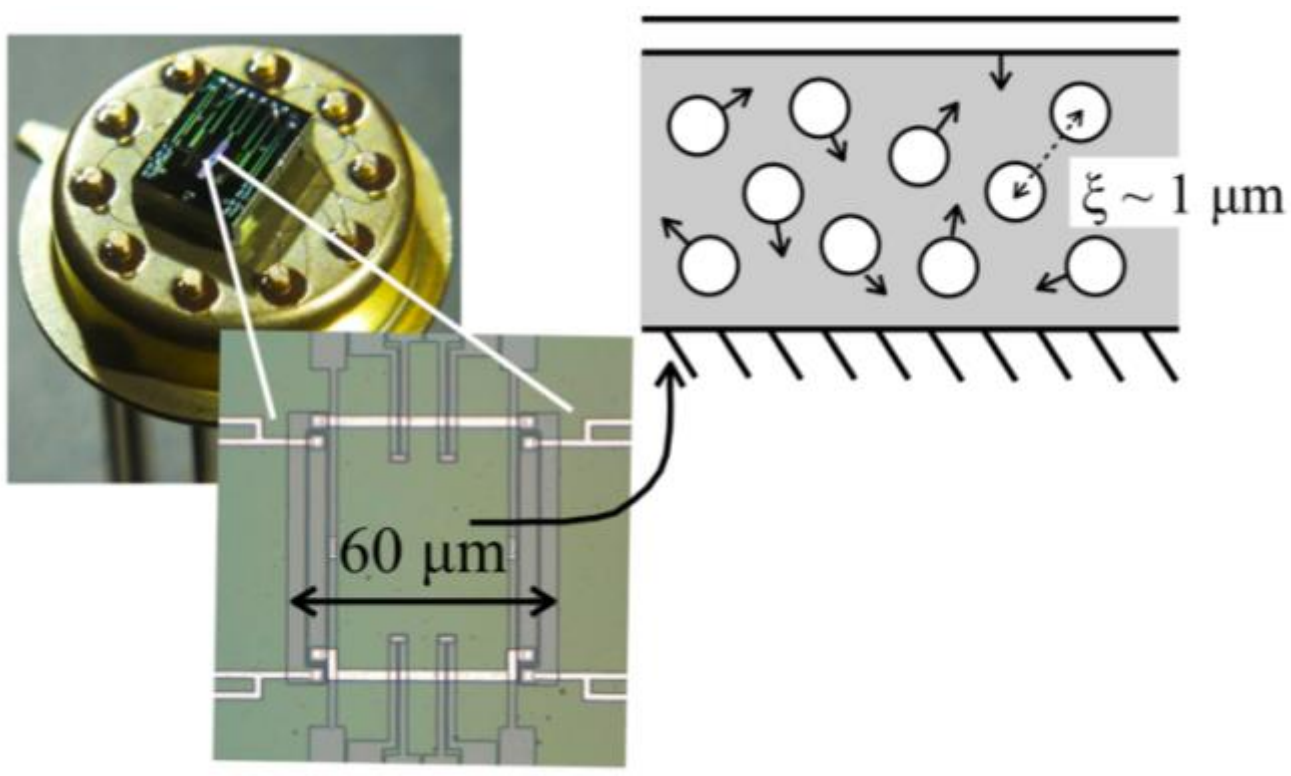

As a supercooled liquid is cooled, molecular motion slows and the system eventually falls out of equilibrium at the glass transition temperature $T_g$. Nanometer length scale processes are thought to control the behavior of supercooled liquids and glasses. For example, the size of cooperatively rearranging regions[1] in supercooled liquids has been estimated to be a few nm. As a second example, dynamics in many supercooled liquids are spatially heterogeneous[2-4] on the scale of 1-5 nm.[5-7] A third nanoscale aspect of supercooled liquid dynamics is self-diffusion. Near $T_g$, a crossover from Fickian to non-Fickian diffusion occurs at a few nm.[8,9] Finally, the glass-forming characteristics of metallic and inorganic glasses are commonly discussed in terms of "medium-range order", yet another important length scale of a few nm.[10,11]

In contrast to these nanoscale processes, here we describe a 1 μm length scale that controls the isothermal transformation of vapor-deposited glasses of indomethacin (IMC) into the supercooled liquid. Glasses prepared by physical vapor deposition can have highly efficient amorphous packing and reside low on the potential energy landscape.[12-15] Previous experiments established that slow depositions onto substrates near 0.85 $T_g$ produce high density, low enthalpy, low energy, and high stability glasses.[12,13,16] Here we use quasi-isothermal nanocalorimetry to observe the transformation of low energy glasses into the equilibrium supercooled liquid during annealing above the conventional $T_g$. As the film thickness increases from 75 to 600 nm, the transformation time increases by more than an order of magnitude, consistent with a surface-initiated transformation mechanism. In contrast, for thicknesses greater than 1 μm, a bulk process dominates. The observation of size-dependent transformation kinetics for glass samples approaching 1 μm is unprecedented. We interpret these results to indicate that a 1 μm length scale describes the average distance between transformation sites in the low energy glass.

Figure 1 shows the reversing heat capacity $C_p$ for an as-deposited stable IMC glass and the post-anneal sample. Samples were vapor-deposisted directly onto nanocalorimeters and the reversing Cp (at 20 Hz) was measured from the temperature oscillation (~0.25 K) resulting from an applied AC voltage. The as-deposited glass was heated at about 5 K/min from 298 K to the annealing temperature $T_{anneal}$ of 320 K [$T_g$ = 315 K (10 K/min DSC)]. The sample was then held isothermally until $C_p$ reached a constant value, indicating the complete transformation of the stable as-deposited glass as shown in the inset of Figure 1; this step is actually quasi-isothermal since the temperature of the nanocalorimeter oscillates slightly. After annealing, the sample was cooled and then heating/cooling scans were performed at a rate of 1 K/min to obtain the reversing $C_p$ of the ordinary glass and supercooled liquid as a function of temperature. The behavior of the annealed samples is consistent with previous AC calorimetry experiments on IMC[17] and other supercooled liquids;[18] at low temperatures, only vibrations can respond to the small temperature oscillation while molecular rearrangements also contribute to $C_p$ at high temperatures. Even though $C_p$ at the end of isothermal annealing is similar to $C_p$ for the ordinary glass, from a thermodynamic perspective the sample is a supercooled liquid showing no hysteresis upon subsequent heating and cooling.

A striking feature of Figure 1 is the significant difference in $C_p$ between the as-deposited and post-anneal glass. The difference in $C_p$ is unexpectedly large and is estimated to be 16 ± 8 J/(mol K). In comparison, Chang, Bestul, and coworkers measured the difference in $C_p$ between aged and liquid quenched glasses and found differences of about 1 J/(mol K) for a number of different systems.[19-22] Near $T_g$, the known difference in $C_p$ between the ordinary glass and γ crystalline polymorph of IMC is 13 J/(mol K).[23] The considerable difference in $C_p$ between as-deposited and annealed samples allows us to observe the isothermal evolution of the low energy glass illustrated by the inset of Figure 1.

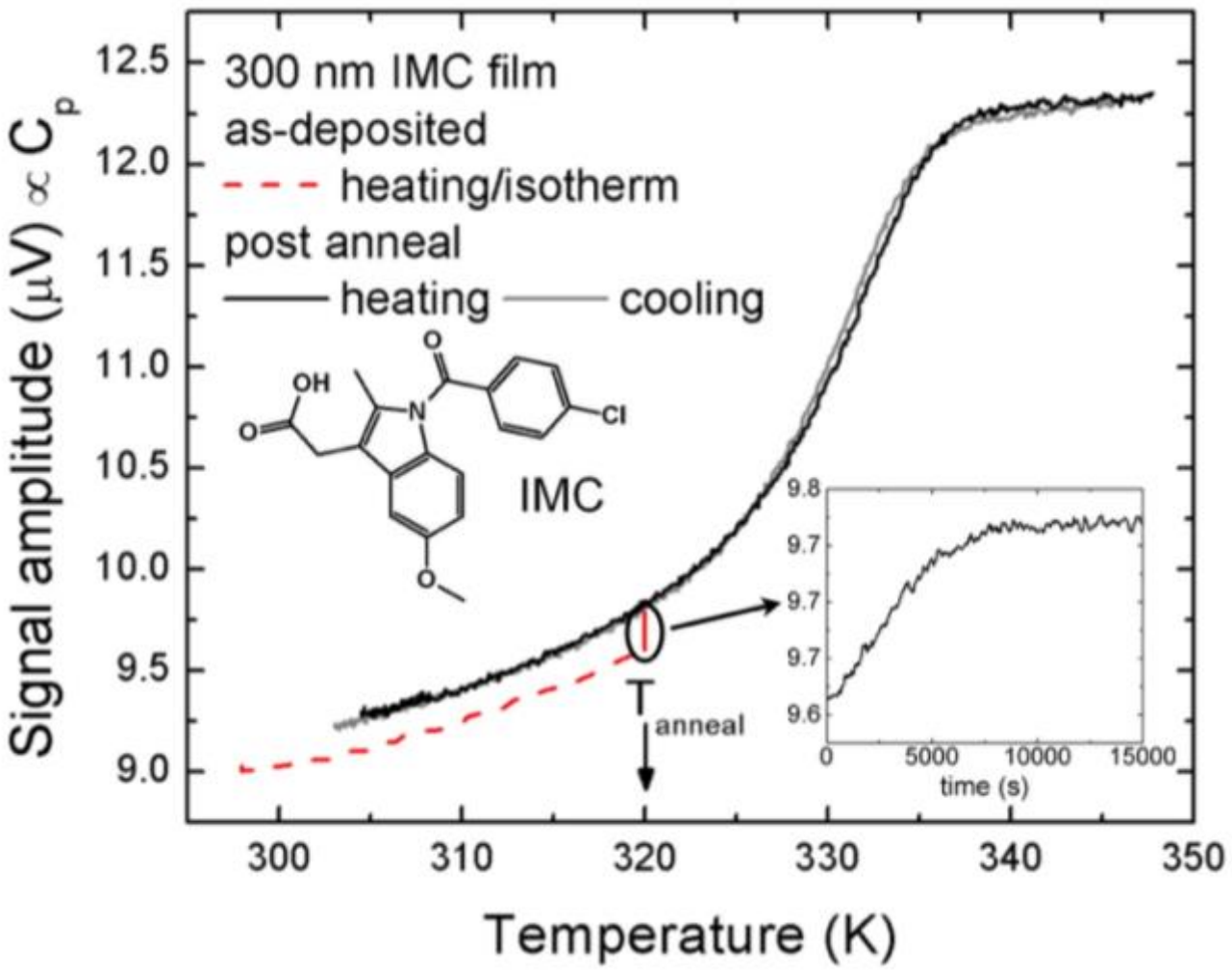

Figure 1: The signal amplitude, which is proportional to heat capacity $C_p$ at 20 Hz, for an as-deposited and post-anneal 300 nm indomethacin glass film. The dashed red curve represents the as-deposited glass. An isothermal anneal at 320 K ($T_g$(DSC) = 315 K) transforms the glass into the supercooled liquid (inset). Once the thermal history is erased, the sample is heated (black) and cooled (gray) between 305 K to 343 K. The similarity between the heating and cooling curves confirms that annealing completely transforms the initial low energy glass.

Figure 2 shows the isothermal transformation kinetics for stable, low energy IMC glasses of five different thicknesses. $\phi_{SG}$ describes the fraction of the sample with the heat capacity of the stable glass. Because irreversible heat flow (e.g., enthalpy relaxation) is excluded from our signal, $\phi_{SG}$ can be directly calculated from $C_p$. Figure 2 shows that the time needed to transform the entire film increases with sample thickness. This dependence is contrary to the conventional view that sample size is irrelevant for glass transformation kinetics.[24-26] For films that are 300 nm or less, a nearly linear transformation takes place. This is consistent with a surface-initiated growth process in which the stable, low energy glass transformation starts at the air interface and proceeds into the bulk at a constant rate leaving supercooled liquid behind. Similar behavior was recently reported for vapor-deposited glasses of tris-naphthylbenzene.[27] For thicker films, the non-linear shapes of $\phi_{SG}$ indicate that the transformation mechanism has changed. In some cases $\phi_{SG}$ becomes negative due to slow instrumental drift during the isothermal anneal; this was confirmed by experiments performed on two empty sensors.

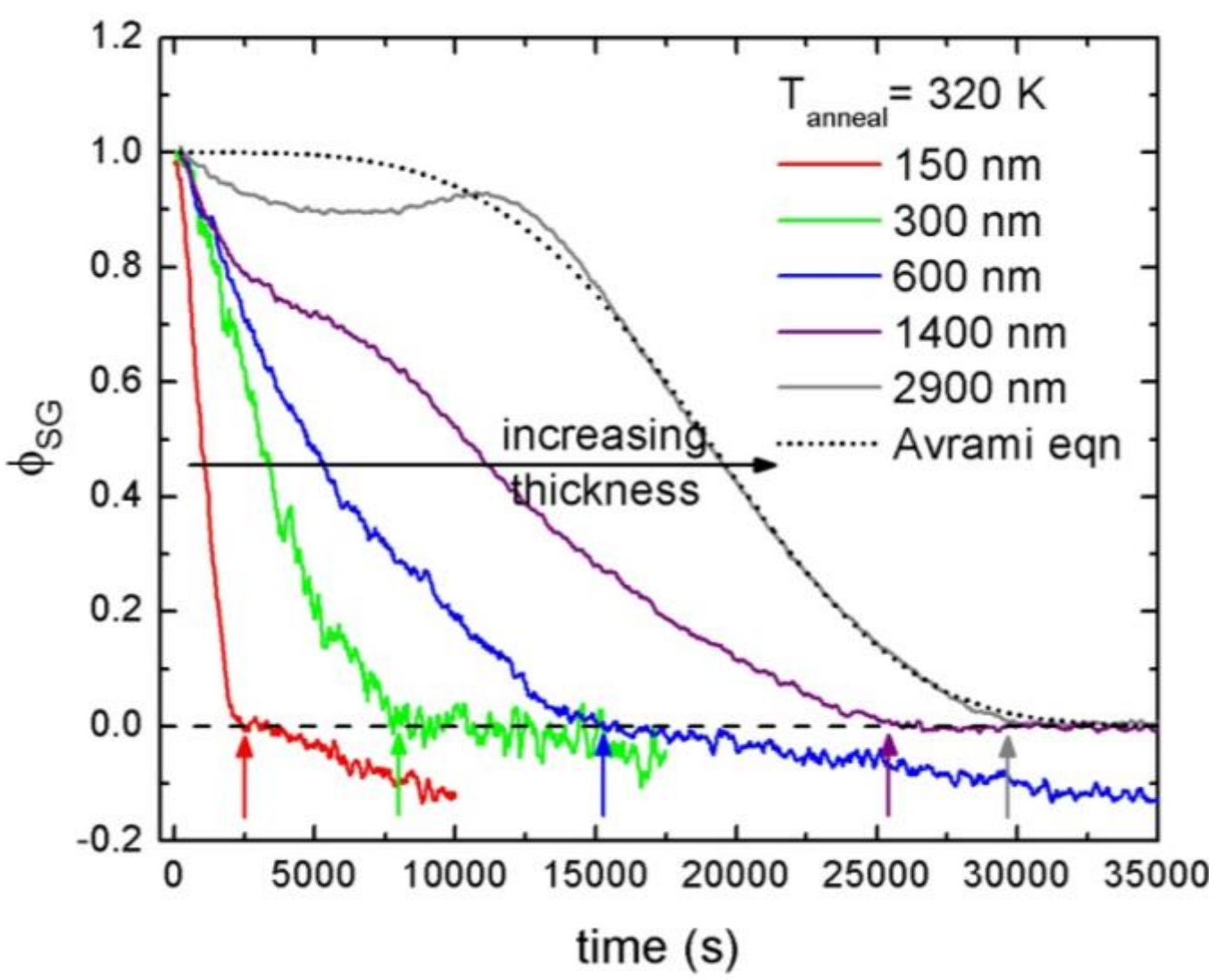

Figure 2: Isothermal transformation of stable, low energy IMC glasses of various thicknesses; $\phi_{SG}$ is the fraction of the sample with the heat capacity of the stable glass. Vertical arrows indicate the time to completely transform films of different thicknesses. For the thickest film, the Avrami equation [$\phi_{SG} = \exp(-Kt^m)$] was fit to the data yielding $K = 4.68 \times 10^{-17}$ and $m = 3.78$.

Figure 3 summarizes the transformation times obtained for low energy IMC glasses via nanocalorimetry as a function of $T_{anneal}$ and thickness. Similar isothermal calorimetry experiments performed on 30 μm samples using conventional DSC are also shown.[17]

Samples thinner than 600 nm (filled symbols) show a clear dependence on sample thickness. Even the thinnest low energy glasses transform slowly in comparison with the structural relaxation time $\tau_\alpha$ of supercooled IMC;[28] conventionally aged glasses typically require less than 10 $\tau_\alpha$ to thermally rejuvenate.[29] All films, with the possible exception of the 75 nm samples, show the same dependence of the transformation time upon annealing temperature when experimental errors are considered.

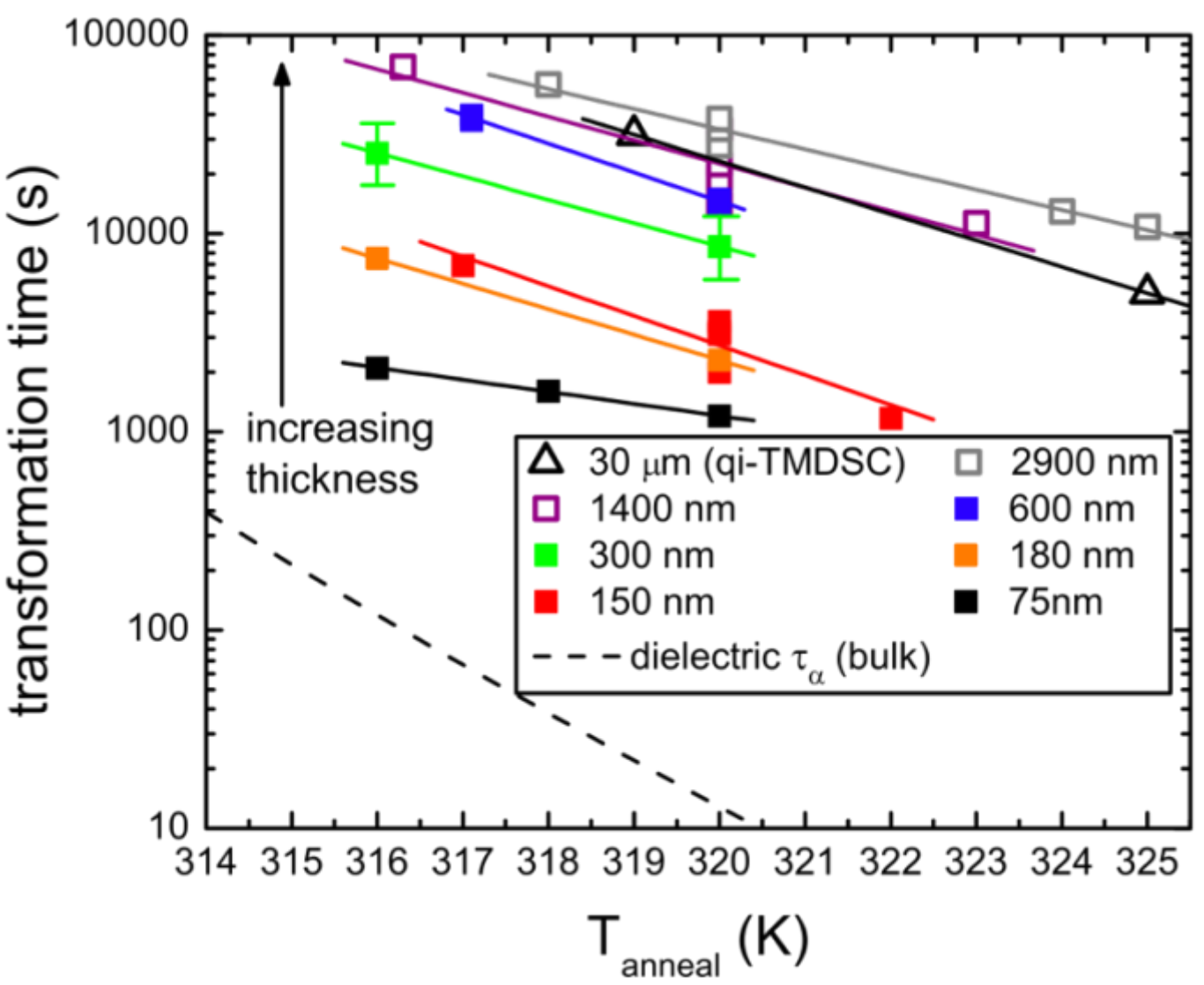

Figure 3: Transformation times for low energy IMC glasses of various thicknesses as a function of annealing temperature. Squares are independent nanocalorimetry experiments with lines of best fit through the data; representative error bars are shown. qi-TMDSC measurements on 30 μm samples (triangles) are also shown.[17] The dashed line describes the bulk dielectric relaxation times for supercooled IMC liquid.[28]

Figure 4 clearly displays two thickness regimes for the isothermal transformation of low energy IMC glasses and allows for the determination of the surface-initiated growth front velocity. For film thicknesses of 600 nm or less, a line of slope one is fit through the data. This is based on the picture of a single growth front moving into the stable glass from the free surface at a constant rate. Such a surface-initiated growth front has been observed in preliminary SIMS experiments on IMC stable glasses; the front velocity observed in those experiments is consistent with the value of 0.04 nm/s deduced from Figure 4 for $T_{anneal} = 320$ K. A second regime, in which the transformation time is independent of sample thickness, is established for films between 1.4 and 30 μm.

We rationalize the results shown in Figure 4 by a crossover in the dominant transformation process of the glass as a function of film thickness. For the thinnest films, the packing is so efficient that no molecular rearrangement occurs in the interior of the sample before the surface-initiated growth front has moved through the film. For the thickest films, the transformation into the supercooled liquid occurs in the bulk of the sample before the front has propagated through the entire sample. For intermediate film thickness, both surface-initiated and bulk transformation events contribute significantly to the overall transformation kinetics.

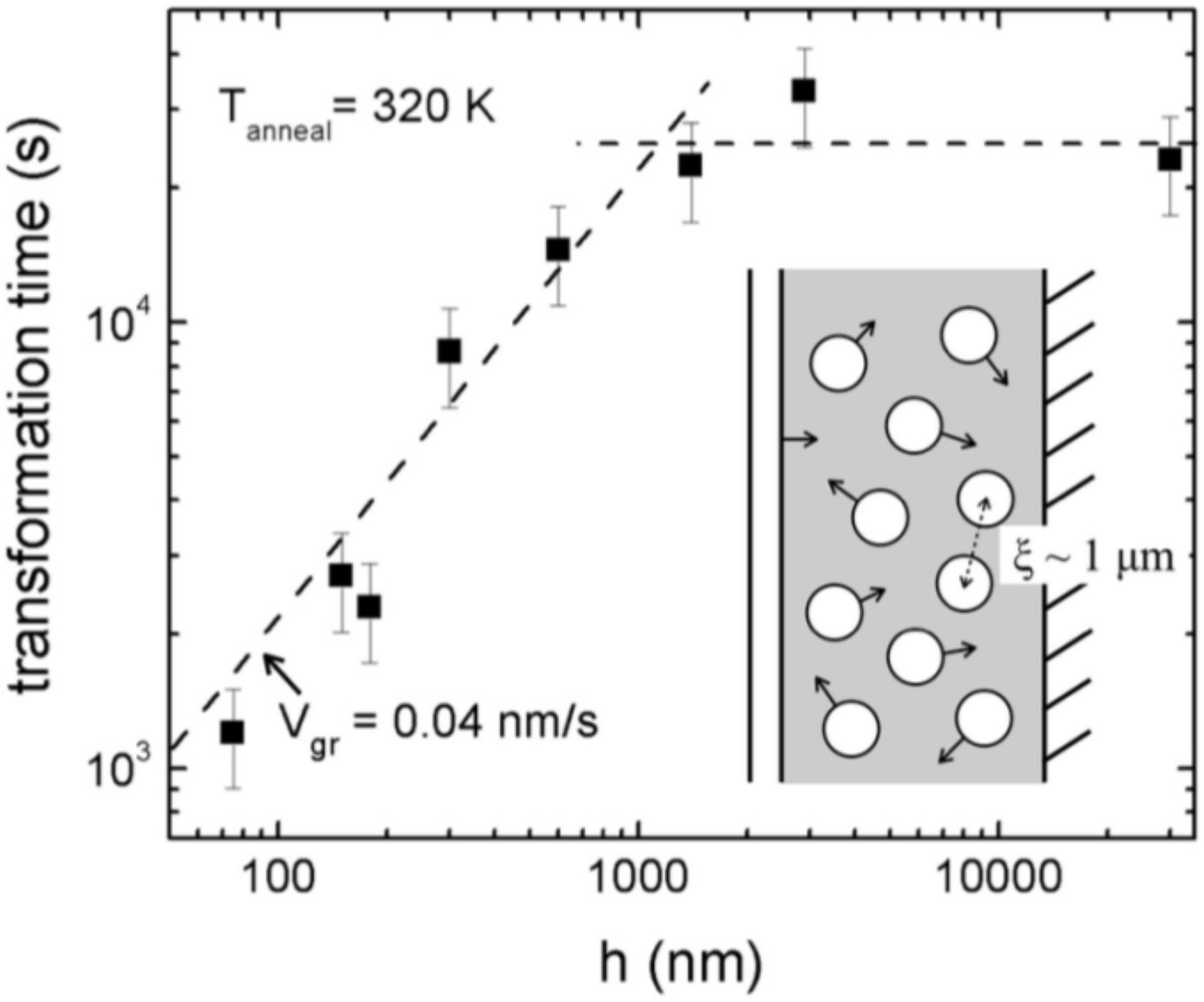

Figure 4: Crossover from thickness-dependent to thickness-independent stable glass transformation times at $T_{anneal}$ = 320 K. A line of best fit with slope = 1 is drawn through the thin film data while a line with slope = 0 is drawn for thicker films. Inset: Schematic representation of the surface and bulk transformation pathways. In each case, the supercooled liquid (white) grows into the stable glass (gray). The free surface is on the left.

The intersection of the dashed lines in Figure 4 clearly shows that 1 μm is the crossover thickness between the surface-initiated and bulk transformation mechanisms. We take our analysis one step further and interpret this crossover thickness as indicating a structural length scale in the low energy glass. As described below, we interpret this length as the average distance between transformation initiation sites in the bulk of the low energy glass.

Recently, Wolynes used the random first order transition theory (RFOT) to describe the thermal rejuvenation of aged glasses,[30] i.e., the evolution of the aged glass into the supercooled liquid following a temperature jump from below $T_g$. Reference 30 describes radially propagating fronts in the bulk that originate in high mobility regions of the glass. Adopting this perspective, we describe the transformation kinetics by the equation $\xi \approx v\, t_R$; the transformation is complete when growing regions of supercooled liquid impinge upon each other. A length scale ($\xi$), the total transformation time ($t_R$), and the front velocity ($v$) are needed to describe this process.

The application of this bulk transformation mechanism to the data shown in Figure 4 leads to the conclusion that the sites that initiate transformation in low energy IMC glasses are about 1 μm apart. For this calculation, we take $t_R$ to be the transformation time of the thickest films (25,000 seconds) and assume that the growth front velocity in the bulk (0.04 nm/s) is the same as it is near the free surface. The inset of Figure 4 provides a schematic representation of the surface-initiated and bulk growth processes for these low energy glasses. While we expect that both processes occur simultaneously in all samples, the transformation kinetics of thick films will naturally be dominated by the bulk mechanism.

Three arguments support our interpretation of the low energy glass transformation kinetics in terms of a 1 μm length scale between sites that initiate the transformation to the supercooled liquid. First, roughly the same length scale is deduced from the data in Figure 3 at other annealing temperatures, as expected for a *structural* length scale. Second, Brillouin light scattering experiments on 10 μm low energy IMC glasses display evidence for sample heterogeneity on a length scale greater than 500 nm ($> 1/q_{scattering}$) when $\phi_{SG} \approx 0.5$.[31] This result, deduced from changes in the lineshapes of the scattering peaks, must be observed if our proposed mechanism is correct. Finally, the transformation kinetics of our thickest samples can reasonably be fit using the Avrami equation, as shown in Figure 2. The Avrami equation has a characteristic sigmoidal shape and is often used to describe the nucleation and growth of crystals from a supercooled liquid.[32] It is also well-suited to describe the growth of the supercooled liquid "bubbles"[33] shown in the inset to Figure 4. Initially, the predicted transformation kinetics are slow because the radii of the growing objects (regions of supercooled liquid) are small. If the radial growth velocity is constant, the transformation then accelerates due to the cubic relation between the radius and the volume. Eventually the transformation slows when the growing regions of supercooled liquid begin to coalesce. While the reasonable fit of our thick film data to the Avrami equation supports our proposed transformation mechanism, this model does not explain the small increase in $\phi_{SG}$ (at about 10000 s) for the 2900 nm film in Figure 2. Further investigation of this interesting feature by techniques such as ellipsometry is warranted.

Two different physical pictures can be invoked to explain the initiation of the bulk transformation. Using the language of crystal nucleation, the initiation may take place either homogeneously or heterogeneously. For "heterogeneous nucleation", we imagine a well-packed low energy glass containing a very low concentration of "defects" that are quite mobile as a result of less efficient amorphous packing. Upon jumping the temperature above $T_g$, these small regions, which are separated by ~1 μm, are able to rearrange and begin the transformation to the supercooled liquid. (In this picture, the "defects" are structural not compositional.) In the analogous homogeneous nucleation view, conversely, the 1μm length scale is a simple consequence of the growth rate and the rate at which glassy regions become mobile (similar to the nucleation rate), i.e., any site in the glass has an extremely small probability of starting a local transformation to the supercooled liquid. Although we cannot distinguish between these two possibilities, the Avrami exponent (~ 4) is consistent with the sporadic nucleation of the "homogeneous" view.

The results presented here provide new insight into the nature of molecular packing in glasses. We have shown that a 1 μm length scale controls the thermal stability of low energy glasses of indomethacin. We interpret this length scale as describing the average distance between mobility initiation sites in the bulk glass. When the sample is heated above $T_g$, regions of supercooled liquid grow from these sites allowing the entire sample to be transformed. Given the length scale and structural nature of the transformation process, light scattering experiments could provide further information; these experiments will be challenging due to the small change in the index of refraction between the stable and ordinary glass. Although we expect that the transformation mechanism described here applies to all low energy glasses independent of preparation method, 1 μm need not be an upper limit for the distance between initiation sites. Even lower energy glasses[12,13,16] would be expected[30] to require longer transformation times and the distance between transformation initiation sites could potentially reach 10 μm or more.

Acknowledgements:

We gratefully acknowledge joint funding from the U.S. National Science Foundation and the German Science Foundation (DFG-SCHI 331 14-1, CHE-0724062 and CHE-0605136). We thank Peter Wolynes for helpful conversations.

Experimental:

IMC films with thicknesses varying from 75 nm to 3 μm were prepared by physical vapor deposition and analyzed with AC nanocalorimetry[34] techniques. Physical vapor deposition was carried out as described previously.[12,13,16] The rate of deposition for all films described here was 0.2 ± 0.03 nm/s and was monitored with a quartz crystal microbalance. Based upon comparisons with ellipsometry and X-ray reflectivity, the absolute thicknesses reported here are accurate within 10% for films less than 1 μm. IMC was deposited directly onto the nanocalorimeters (Xensor Integrations, XEN-39321), which were held at a temperature of 265 K (0.85 $T_g$) during the deposition. The active area of the nanocalorimeter is 60 μm x 60 μm.

AC nanocalorimetry was performed *ex situ,* in an apparatus similar to that described previously,[35] to measure the reversing heat capacity $C_p$ at a frequency of 20 Hz. A small temperature oscillation results from an AC voltage applied to the nanocalorimeter heater by a lock-in amplifier. The temperature is measured by an integrated thermopile on the nanocalorimeter chip. The reversing $C_p$ of the IMC film is directly proportional to the differential temperature amplitude between reference and sample nanocalorimeters.[35] The reference and sample nanocalorimeters were placed in a temperature controlled housing to allow temperature jumps and linear heating/cooling ramps. Each nanocalorimetry run was performed in a dry nitrogen gas atmosphere to eliminate the effect of water on the sample.[36,37]